\documentclass[10pt]{article}

\usepackage{amsmath}
\usepackage{amssymb}
\usepackage{bm}
\usepackage{graphicx,color}
\usepackage{slashed}
\usepackage{mathrsfs}
\usepackage{authblk}
\usepackage{amsmath}
\usepackage{amssymb}
\usepackage{bm}
\usepackage{graphicx,color}
\usepackage{slashed}
\usepackage{subfigure}
\usepackage{float}
\usepackage{cite}

\date{}

\title{\bf Complex-valued Holographic Pseudo Entropy via Real-time AdS/CFT Correspondence}

\author[1]{Zhou Chen\footnote{chenzh339@mail2.sysu.edu.cn}}
\affil[1]{Department of Physics, Southern University of Science and Technology, Shenzhen 518055, China}

\begin{document}
\maketitle

\begin{abstract}
The pseudo entropy is a promising recent generalization of the entanglement entropy to the situations in which both the initial and final state are involved, with the density matrix promoted to the transition matrix. However, in contrast to the non-Hermiticity of the generic transition matrix, the holographic pseudo entropy formulated via the Euclidean AdS/CFT turns out to be always real-valued, which potentially conceals the crucial natures of this novel quantity. In this note, we make the first attempt to formulate a real-time prescription for computations to incorporate naturally the pseudo entropy, as a generally complex-valued entanglement measure, into the AdS/CFT context. It is then conjectured that the holographic pseudo entropy is dual to the extremal codimension-2 area surface in the generally time-dependent Lorentzian asymptotically AdS spacetime, but may also receive imaginary contribution from the regularized extrinsic curvature term of the area surface, which is not included in the covariant holographic entanglement entropy. In this real-time prescription, the holographic pseudo entropy can be considered as a generalization of the covariant holographic entanglement entropy, as well.
\end{abstract}

\newpage

\section{Introduction}
The AdS/CFT correspondence establishes an equivalence between $(d+1)$-dimensional gravity theory in asymptotically AdS spacetime and $d$-dimensional conformal field theory on the boundary\cite{Maldacena:1997re,Gubser:1998bc,Witten:1998qj}, and has motivated much research related to quantum information theory in the high-energy physics in recent years. One of the most important aspect of AdS/CFT correspondence is the geometrization of quantum entanglement, which provides a geometric prescription to compute entanglement entropy in holographic field theory.

This geometric prescription was first given by Ryu and Takayanagi (RT) in \cite{Ryu:2006bv,Ryu:2006ef} for the static Lorentzian asymptotically AdS spacetime, and was subsequently generalized to generic time-dependent situations by Hubeny, Rangamani, and Takayanagi (HRT) in \cite{Hubeny:2007xt}. The RT/HRT proposals tell us that the entanglement entropy of boundary CFT is dual to the area of the bulk codimension-2 extremal surface which is homologous to the boundary entangling surfaces. In a static asymptotically AdS spacetime, the extremal surface sits on the canonical time slice and coincides with the minimal area surface, while the Wick rotation into the Euclidean AdS/CFT setup is straightforward. However, such the Wick rotation becomes ambiguous for time-dependent asymptotically AdS spacetime, raising a question on the physical meanings of the area of the minimal surfaces in the time-dependent Euclidean asymptotically AdS space. Motivated by this, \cite{PhysRevD.103.026005} introduced a novel quantity, \emph{viz}., the pseudo entropy, as a generalization of the entanglement entropy, which is just dual to the area of the minimal surfaces in the time-dependent Euclidean asymptotically AdS space. However, the holographic pseudo entropy (HPE) formulated in Euclidean AdS space is always real-valued, provided that the bulk metric is real-valued, although the generic transition matrix, from which the pseudo entropy is constructed, is non-Hermitian by definition. Recently, the pseudo entropy in the dS/CFT duality and/or time-like entanglement entropy are also studied in \cite{Doi:2022iyj,Doi:2023zaf,Li:2022tsv,Narayan:2022afv} (see also \cite{Wang:2018jva} for earlier consideration), which are both complex-valued and are argued to be correctly understood as pseudo entropy \cite{Doi:2022iyj}. For more developments of pseudo entropy, refer to \cite{PhysRevLett.126.081601, PhysRevResearch.3.033254, Camilo:2021dtt, PhysRevD.104.L121902, Miyaji:2021lcq, PhysRevD.105.126026, Mori:2022xec, Akal:2022qei, Berkooz:2022fso, Mukherjee:2022jac, Guo:2022sfl, Miyaji:2022dna, Akal:2021dqt, Bhattacharya:2022wlp, Guo:2022jzs, He:2023eap} for examples.

In spite of the original motivation, it can be rational to suspect that the pseudo entropy as a generalized entanglement measure is not inherently a CFT quantity which has its dual only to the Euclidean asymptotically AdS space. One may also expect its complex \emph{nature} to manifest itself via AdS/CFT duality in a somewhat intuitive manner, as via the non-Hermiticity of transition matrix he/she immediately deduces. Such an expectation may come from roughly a two-fold reason: for the sake of computational purpose, one may want to figure out an appropriate prescription to incorporate the generic \emph{complex} HPE as new measure of entanglement for future use; while such a prescription may in turn help one understand better its \emph{complex} nature in the AdS/CFT context. Motivated by these, instead of working with the Euclidean settings as initiated in the previous literature, we provide in this note a Lorentzian or real-time path integral prescription for the computations of HPE, which shows that the HPE generally takes complex values. Using this real-time prescription, we argue that the HPE may be dual to the extremal codimension-2 area surface in the (generally time-dependent) Lorentzian asymptotically AdS spacetime, and receive imaginary contribution from the regularized extrinsic curvature term in addition to the area of the extremal surface. In this sense, we also argue that the HPE can be considered as a generalization of the covariant holographic entanglement entropy that is just dual to the area of the extremal codimension-2 surface.

In Section 2, we provide the basic real-time path integral presentation for pseudo entropy in the QFT realm. In Section 3, we recast the previous real-time prescription into the Schwinger-Keldysh formalism and compare with that of the covariant holographic entanglement entropy. In section 4, the complex nature of the holographic pseudo entropy and its dual in the bulk are discussed.
\section{Real-time prescription}
The pseudo entropy is a generalization of von Neumann entanglement entropy with the density matrix replaced by the transition matrices\cite{PhysRevD.103.026005},
\begin{equation}\label{transition matrix}
	\tau^{i\vert j}=\frac{\vert \Psi_i\rangle\langle \Psi_j\vert}{\langle\Psi_j\vert\Psi_i\rangle},
\end{equation}
where $i,j=1,2$ (we let $i\neq j$), $\vert\Psi_1\rangle$ and $\vert\Psi_2\rangle$ are two pure quantum states which are non-orthogonal: $\langle \Psi_1\vert \Psi_2\rangle\neq 0 $. Taking the limit $\vert \Psi_2\rangle =\vert \Psi_1\rangle$, the transition matrices reduces to the ordinary density matrix. Dividing the whole system into $\mathcal{A}$ and its complement $\mathcal{A}_c$, the pseudo entropy of $\mathcal{A}$ is defined as
\begin{equation}
S(\tau^{i\vert j}_{\mathcal{A}})=-\textrm{tr}\left[\tau^{i\vert j}_{\mathcal{A}}\log\tau^{i\vert j}_\mathcal{A}\right],
\end{equation}
where $\tau^{i\vert j}_\mathcal{A}=\textrm{tr}_{\mathcal{A}_c}[\tau^{i\vert j}]$ is the reduced transition matrices. Note that the transition matrix is non-Hermitian by definition, such that the pseudo entropy develop complex values generally. For future purpose, we also use a re-expression of the transition matrix \cite{Guo:2022jzs}, \emph{viz.},
\begin{equation}\label{sym tra mat}
	\tau^{i\vert j}=\frac{\rho_i\rho_j}{\textrm{tr}[\rho_i \rho_j]},
\end{equation}
with $\rho_i=\vert\Psi_i\rangle\langle \Psi_i\vert$ the ordinary density matrices.
In this section, we restrict ourselves into the real-time path integral representation of the reduced transition matrices $\tau^{i\vert j}_{\mathcal{A}}$ and pseudo entropy $S(\tau^{i\vert j}_\mathcal{A})$, in the QFT realm. Below, we choose the $t=0$ Cauchy slice $\Sigma_0\equiv\mathcal{A}\cup\mathcal{A}_c$ to compute these quantities.

Let us proceed with the transition amplitude $\langle \Psi_j\vert \Psi_i\rangle$, since a generic reduced transition matrix can be simply obtained by cutting open the path integral of this amplitude along some specified regions on a Cauchy slice, up to the conventional normalization. We place the initial state $\vert \Psi_1\rangle$ in the past, with the final state $\vert \Psi_2\rangle$ in the future. Doing QFT, we recast the amplitude into a path integral, \emph{e.g.} write down,
\begin{equation}\label{2.1}
\langle \Psi_j \vert \Psi_i\rangle=\int \mathcal{D}\phi_0\langle \Psi_j\vert \phi_0\rangle\langle \phi_0\vert \Psi_i\rangle,
\end{equation}
where we use a short hand $\phi(t=0,\vec x) \equiv \phi_0$ where the spatial coordinates $\vec x$ are suppressed for the clarity. In \eqref{2.1} one effectively cuts open the path integral along $\Sigma_{0}$ by inserting corresponding instantaneous states of interest that characterize the system, and then glue it up again throughout $\Sigma_{0}$. On the other hands, viewing \eqref{2.1} as the total trace of $\vert \Psi_i\rangle\langle \Psi_j\vert$, the reduced transition matrices $\tau^{i\vert j}_\mathcal{A}$ can thus be similarly obtained by a partial gluing, leaving the cut along the subregion $\mathcal{A}$ on the Cauchy slice $\Sigma_0$ remain open.

The path integral \eqref{2.1} can be formulated in the Euclidean setup, which however leads to the always real-valued holographic pseudo entropy. The idea in this note is to deform the path-integral time-contour from the Euclidean one to a piece-wise extension with a mixture of Euclidean and Lorentzian path integrals\cite{Skenderis:2008dh,vanRees:2009rw}. In such piece-wise prescription and for our purpose, the Euclidean segments serve to produce the initial and final wavefunctions (or conditions) of interest, whereas the non-trivial evolution is encoded into the Lorentzian ones. We do not presume here the Wick rotation from the Lorentzian to the Euclidean, leaving it as the future problem to examine such a rotation. We decompose the wavefunctionals as below,
\begin{equation}\label{2.2}
\begin{split}
 \langle \phi_0\vert \Psi_1\rangle&=\int\mathcal{D}\phi_{-} \langle \phi_0\vert\phi_-\rangle\langle\phi_-\vert  \Psi_1\rangle,\\
   \langle \Psi_2\vert \phi_0\rangle&=\int\mathcal{D}\phi_{+} \langle \Psi_2\vert\phi_+\rangle\langle\phi_+\vert  \phi_0\rangle,
\end{split}
\end{equation}
\begin{figure}[ht]
	\centering
	\includegraphics[width=0.55\linewidth]{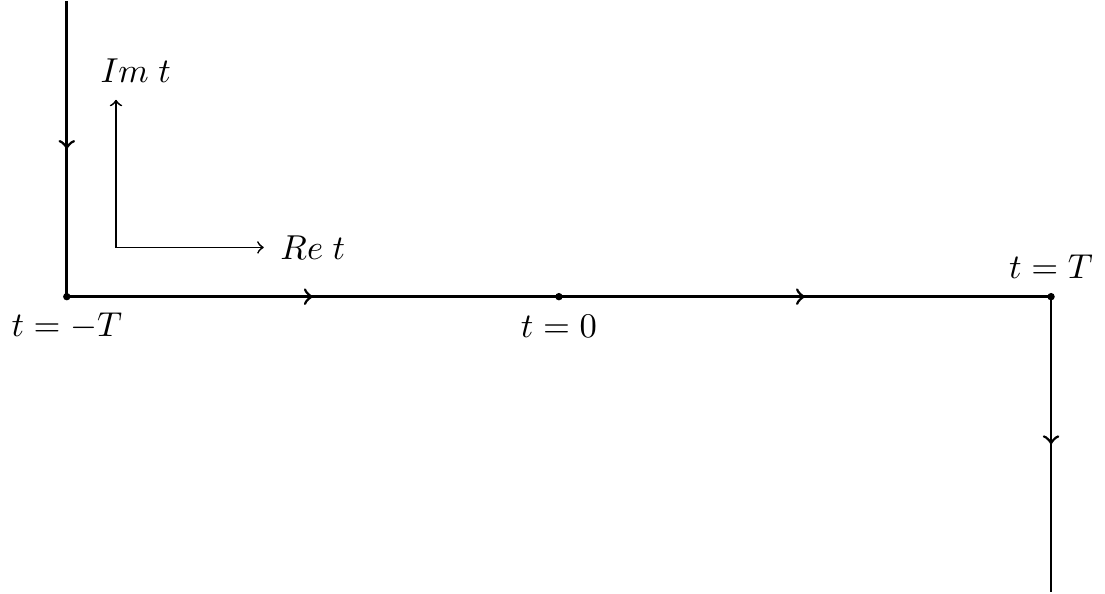}
	\caption{\footnotesize Complex-time contour for $\langle \Psi_2\vert \Psi_1\rangle$ with the direction of time flow shown explicitly and spatial coordinates expressed. For $\langle \Psi_1\vert \Psi_2\rangle$, one inverses the flow of time in the Lorentzian segment and translates the Euclidean segments vertically. For our purpose, the Hamiltonian in the two Euclidean segments are set to be time-independent, and different from each other so that the initial and final state thus produced differ, which finally ends up with a non-trivial time-dependence in the Lorentzian segment violating the time-translation and -reflection symmetry.}\label{comptim}
\end{figure}
where the subscriptions `$-$, $0$, $+$' indicate that integrated field configurations lives on the Cauchy slices $\Sigma_{-T}$, $\Sigma_{0}$ and $\Sigma_{+T}$, respectively. Hereafter, we set $T>0$ to fix the time direction. The strategy now is to arrange $\langle \phi_0\vert\phi_\pm\rangle$ into the real-time axes, and think of $\langle\phi_\mp\vert\Psi\rangle$ as the static Euclidean path integrals which prepare the initial/final state on $\Sigma_{\mp T}$. This piece-wise arrangement leads to a complex-time contour as depicted in figure \ref{comptim}. Note that the path integral contour of $\langle\Psi_1\vert\Psi_2\rangle$ differs from that of $\langle\Psi_2\vert\Psi_1\rangle$ by a time reversal, since we fix $\vert\Psi_1\rangle$ as the initial state and $\vert\Psi_2\rangle$ as the final state in the real-time evolution sense. This time-reversal feature can be formulated to be exact, noting the real-time evolution shall be prescribed to be unitary. We refer to the spacetimes related to $\langle\Psi_2\vert\Psi_1\rangle$ and $\langle\Psi_1\vert\Psi_2\rangle$ as the forward spacetime $\mathcal{B}^{1\vert2}$ and the backward spacetime $\mathcal{B}^{2\vert1}$, respectively. Although the ``evolution'' from $\vert \Psi_1\rangle$ to $\vert\Psi_2\rangle$ is unnecessarily unitary by definition, we can consider only the unitary evolution w.l.o.g, noting that the non-unitarity in the overlap $\langle\Psi_i\vert\Psi_j\rangle$ can be somehow encoded into the Euclidean segments as the state preparation process, leaving the real-time evolution still unitary. Or, one can consider a certain unitary evolution with \emph{truly} the initial and final state, $\vert \Psi_1\rangle$ and $\vert \Psi_2\rangle$. Especially, if $\vert \Psi_1\rangle$ is chosen as a pure ground state, one can set the corresponding Euclidean segment to go from time $-T+i \beta$ to $-T$ along the imaginary-time axis with a large $\beta$, while the final pure ground state wavefunctional of $\vert \Psi_2\rangle$ can be similarly constructed but with a distinct Hamiltonian. Moreover, one can also prepare the excited states from the ground state by inserting various operators into the Euclidean segments as in \cite{PhysRevD.103.026005}, which would be understood by our prescription as deforming the initial or final condition for the real-time solutions. For related considerations on the state preparation, see for example Section 4 in \cite{Colin-Ellerin:2020mva}.
\begin{figure}[ht]
	\centering
	\subfigure[]{\includegraphics[width=0.34\linewidth]{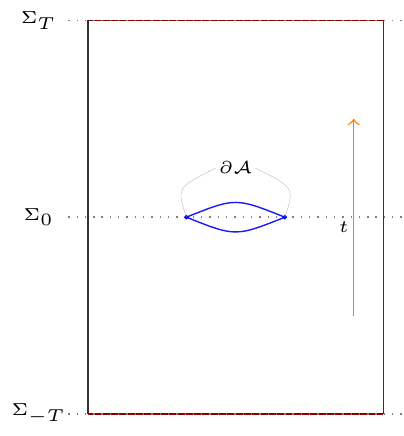}}
	\quad\quad\quad\quad
		\subfigure[]{\includegraphics[width=0.34\linewidth]{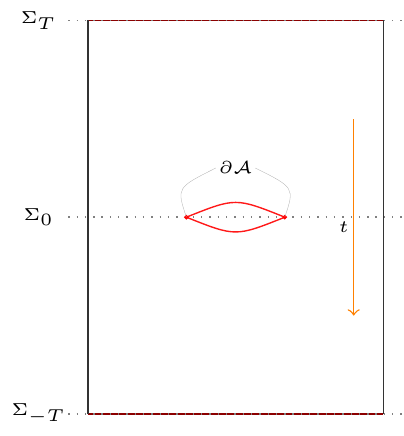}}
	\caption{\footnotesize The path integral representation of (a) $\tau^{1\vert 2}_{\mathcal{A}}$ and (b) $\tau^{2\vert1}_{\mathcal{A}}$, related to the forward spacetime $\mathcal{B}^{1\vert 2}$ and backward spacetime $\mathcal{B}^{2\vert1}$, respectively, with the Euclidean segments omitted. There is a cut along the subregion $\mathcal{A}$ on the Cauchy slice $\Sigma_0=\mathcal{A}\cup\mathcal{A}_c$, with the entangling surface $\partial \mathcal{A}$ explicitly shown. The cuts in different spacetime are shown in different colors, blue or red. The arrows colored in orange indicate the direction of the time flow.} \label{tau12}
\end{figure}

Subsequently, the path integral of $\tau_{\mathcal{A}}^{i\vert j}$ can be obtain by cutting open the path integral of $\langle\Psi_j \vert \Psi_i \rangle$ along the subregion $\mathcal{A}$ on $\Sigma_0$, while they differ from each other by a time reversal, as depicted in figure \ref{tau12}. This time-reversal feature exactly reflects the complex nature of pseudo entropy, and one would find it instructive to consider the $i\epsilon$ prescription which is inherent in the real-time context. For instance, if we begin with preparing the initial/final pure ground state in the distant past/future via an explicit $i\epsilon$ rotating time axes into a complex plane as in the ordinary real-time QFT, instead of slicing open the Euclidean path integrals, we immediately see that the forward and backward time-contours are thus equipped with two ``conjugate rotations'' in the complex-time plane. Requiring this $i\epsilon$ prescription in boundary field theory to generally extend to the bulk geometry via AdS/CFT, we are thus led to the position to double the bulk manifolds in a complex plane and pick up different branches when computing $S(\tau_{\mathcal{A}}^{1\vert 2})$ and $S(\tau_{\mathcal{A}}^{2\vert 1})$. Exactly, such subtleties of the $i\epsilon$ prescription are usually encountered in QFT dealing with various real-time correlation functions,\footnote{Of most relevance to our discussion are perhaps the time-ordered and anti-time-ordered correlation functions.} and of course should be preserved when calculating such correlation functions in real-time AdS/CFT context. We will return to the discussion on this time-reversal feature in the next section, via the Schwinger-Keldysh formalism.

Having the real-time QFT path integral of the reduced transition matrices $\tau^{i\vert j}_{\mathcal{A}}$, we use the replica trick to compute the traces of their $n$th powers, $\textrm{tr}_{\mathcal{A}}(\tau^{i\vert j}_{\mathcal{A}})^{n}$. By gluing cyclically $n$ copies of $\mathcal{B}^{i\vert j}$ across $\mathcal{A}$ with associated boundary condition in a replica $\mathbb{Z}_n$ symmetric manner, we obtain the $n$-fold branched cover spacetime $\mathcal{B}^{i\vert j}_n$ (branched at $\partial\mathcal{A}$) of the original manifold $\mathcal{B}^{i\vert j}$. As usual, there is the fixed point set $\partial\mathcal{A}$ of the replica $\mathbb{Z}_n$ action, on $\mathcal{B}^{i\vert j}_n$. It follows that $\textrm{tr}_{\mathcal{A}}(\tau^{i\vert j}_{\mathcal{A}})^{n}$ are given by the generating functionals $Z[\mathcal{B}^{i\vert j}_n]/Z[\mathcal{B}^{i\vert j}]^n$, where $Z[\mathcal{B}^{i\vert j}]\equiv\langle\Psi_j\vert\Psi_i\rangle$ is the normalization.  Then the $n$th pseudo R\'enyi entropies will be obtained from the standard formula\cite{PhysRevD.103.026005}, viz.,
\begin{equation}\label{renyi1}
	S^{(n)}(\tau^{i\vert j}_{\mathcal{A}})=\frac{1}{1-n}\log \textrm{tr}_{\mathcal{A}}\left(\tau^{i\vert j}_{\mathcal{A}}\right)^{n}=\frac{1}{n-1}\left(I^{i\vert j}_n-n I^{i\vert j}_1\right),
\end{equation}
where $I_n^{i\vert j}=-\log Z[\mathcal{B}^{i\vert j}_n]$, while $\mathcal{B}^{i\vert j}_1=\mathcal{B}^{i\vert j}$. The $n\rightarrow 1$ limit of \eqref{renyi1} after the analytical continuation to non-integer $n$ gives the pseudo entropies $S(\tau^{i\vert j}_{\mathcal{A}})=-\textrm{tr}_{\mathcal{A}}[\tau^{i\vert j}_{\mathcal{A}}\log\tau^{i\vert j}_{\mathcal{A}}]$.

\section{The Schwinger-Keldysh formalism}

In this section, we introduce an equivalent \emph{time-folded} prescription for pseudo entropy which gives rise to the more familiar Schwinger-Keldysh path integral prescription, which has been adopted for the derivation of the HRT proposal in \cite{Dong:2016hjy}, see \cite{Rangamani:2016dms} for a review. To begin with, we recall the symmetric re-expression of the transition matrix, \emph{i.e.} \eqref{sym tra mat}. By taking the partial trace over $\mathcal{A}_c$, we obtain the reduced transition matrices of $\mathcal{A}$, \emph{viz}.,
\begin{equation}\label{recall}
\tau_{\mathcal A}^{i\vert j}=\textrm{tr}_{\mathcal{A}_c}\left[\frac{\rho_i\rho_j}{\textrm{tr}[\rho_i \rho_j]}\right]
\end{equation}
The strategy for computing \eqref{recall} here is to formulate the matrix element of the density matrices, $\langle\phi_0\vert\rho_1\vert \phi^\prime_0\rangle$ and $\langle\phi_0\vert\rho_2\vert \phi^\prime_0\rangle$, separately, with the aforementioned piece-wise description, and then sew them together as doing the matrix multiplication.
\begin{figure}[ht]
	\centering
	\subfigure[]{\includegraphics[width=0.45\linewidth]{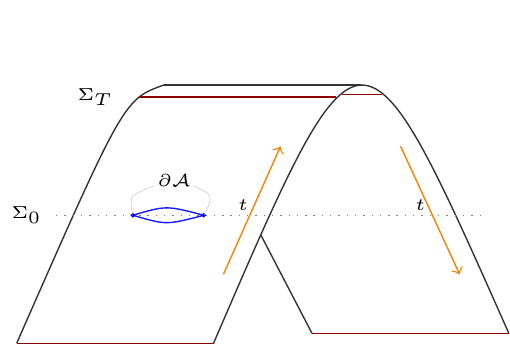}}
	\quad
		\subfigure[]{\includegraphics[width=0.45\linewidth]{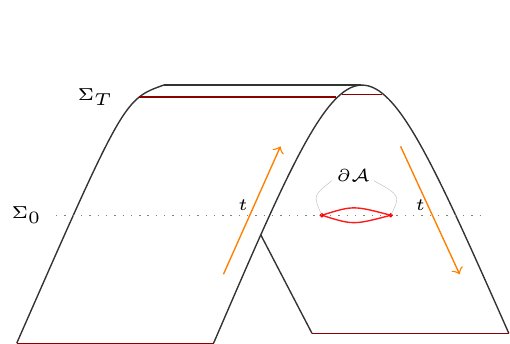}}
	\caption{\footnotesize The time-folded contour of (a) $\tau^{1\vert 2}_\mathcal{A}$ and (b) $\tau^{2\vert1}_{\mathcal{A}}$, which is of the Schwinger-Keldysh type. The path integrals has been glued up on ${\Sigma}_T$ as discussed in the main text. The cut along subregion $\mathcal{A}$ related to $\tau^{1\vert 2}_\mathcal{A}$ or $\tau^{2\vert 1}_\mathcal{A}$ is now arranged into a single branch of the Schwinger-Keldysh contour.}\label{Stau}
\end{figure}
Choosing the conventional Cauchy slice $\Sigma_{0}$ and fixing the initial and final conditions as before, the path integrals for $\tau_{\mathcal A}^{i\vert j}$ are straightforward obtained, as depicted in figure \ref{Stau}. We then immediately recognize the time-folded geometry on which the $\tau_{\mathcal A}^{1\vert 2}$ corresponds to the forward evolution and $\tau_{\mathcal A}^{2\vert 1}$ corresponds to the backward one, with a certain cut along $\mathcal{A}$. Note that the forward and backward Lorentzian segments have been glued on the Cauchy slice $\Sigma_{T}$. To see this, recall that in our piece-wise prescription the two future Euclidean segments prepare the identical final conditions on $\Sigma_{T}$, and each is glued with Lorentzian segment smoothly in a sense that the induced values of the fields and its conjugate momenta are continuous across the gluing surfaces\cite{Skenderis:2008dh,vanRees:2009rw}. In other words, the forward and backward segments have identical boundary conditions on $\Sigma_{T}$, while very single field configuration crosses continuously inward or outward each Euclidean segment. By the virtue of this continuity, provided that the final conditions on $\Sigma_{T}$ is always specified, we can glue up the forward and backward segments across $\Sigma_{T}$ in a sense of time reversal. This formally brings us to the Schwinger-Keldysh contour. Note that, the computation of $\tau_{\mathcal A}^{1\vert 2}$ or $\tau_{\mathcal A}^{2\vert 1}$ involves the forward $\mathcal{B}^{1\vert2}$ or backward $\mathcal{B}^{2\vert1}$ only, with specified initial and final conditions. Thus in the practical situations, one can think of each single branch of the time-folded path integrals as merely imposing the initial or final condition on its companion. This turns out to be in accord with the aforementioned piece-wise construction, modulo some redundancies from the path-integral time-contour deformation. For convenience, we denote the Schwinger-Keldysh geometry introduced above as $\mathcal{B}$, saying $\mathcal{B}$ reduces to $\mathcal{B}^{i\vert j}$ when there is a cut along $\mathcal{A}$ in the forward or backward segment. That is, we only need to compute the pseudo entropy using a single branch of the Schwinger-Keldysh contour, as in figure \ref{tau12}.

It remains to discuss how this time-folded prescription reflects the complex nature of the pseudo entropy. The observation that the path integral of the transition matrices can be recast into the form of the Schwinger-Keldysh implicitly relies on the prerequisite that the initial/final state can be thought of as certain unitary evolution from the immediate states on Cauchy slice $\Sigma_{0}$, \emph{i.e.} $\vert \Psi_2\rangle\sim\mathcal{U}_2\vert \phi_0\rangle$ and $\vert \Psi_1\rangle\sim\mathcal{U}_1^\dagger\vert\phi_0\rangle$, such that the forward and backward segment can be referred to as a conjugate pair of the unitary evolution, saying $\mathcal{U}_2\mathcal{U}_1$ and $\mathcal{U}^\dagger_1\mathcal{U}^\dagger_2$ respectively.\footnote{An appropriate $i\epsilon$ prescription should conform to this conjugate pair of unitary evolution to preserve the causal structure of the various real-time correlation functions. This is implicit in the Schwinger-Keldysh contour.} As a result, the time-reversal feature of our real-time prescription can be treated exactly on the same footing as the Schwinger-Keldysh, which is conventionally defined from $\rho_t=\mathcal{U}_{t,t_i}\rho_{t_i}\mathcal{U}^\dagger_{t,t_i}$. As is known in the Schwinger-Keldysh context, the Schwinger-Keldysh generating functional, $Z\sim e^{i(S^{for}-S^{bac})}$, possesses the CPT-conjugation symmetry, while the forward and backward branches exchange under the CPT conjugation\cite{Haehl:2016pec}.
\begin{figure}[ht]
	\centering
	\includegraphics[width=.6\linewidth]{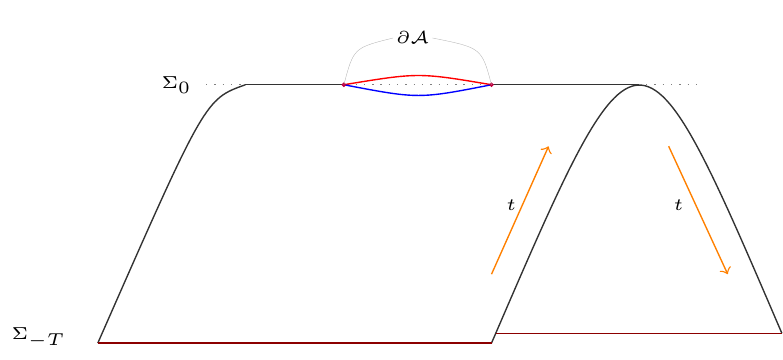}
	\caption{\footnotesize Schwinger-Keldysh construction for $\rho_{\mathcal{A}}(t)$ in the case of the covariant holographic entanglement entropy. $\rho_{\mathcal{A}}(t)$ is the generic time-dependent reduced density matrix. The cut along the subregion $\mathcal{A}$ on the Cauchy slice $\Sigma_0$ of interest lies at the boundary between the forward and backward branch of the Schwinger-Keldysh contour.}\label{Srho}
\end{figure}
It is now crucial to emphasize that the path integral of covariant holographic entanglement entropy (CHEE) enjoys such a CPT-conjugation symmetry as well, which finally renders the R\'enyi and von Neumann entanglement entropies real-valued. To see this, we recall that CHEE is also constructed from a generic time-dependent density matrix, \emph{i.e.}, $\vert \Psi(t)\rangle\langle\Psi(t)\vert=\mathcal{U}_{t,t_i}\vert\Psi(t_i)\rangle\langle\Psi(t_i)\vert\mathcal{U}^\dagger_{t,t_i}$, with $t_i$ being the time(slice) the initial state $\vert\Psi(t_i)\rangle$ is prepared, and $t$ the time the entropy would be computed. We hereafter set $t_i=-T$ and $t=0$ as the convention. Generally speaking, CHEE involves only the causal past of the Cauchy slice $\Sigma_{0}$ of interest, in a fashion of the forward-backward or ket-bra evolution in which one evolves the initial condition on $\Sigma_{-T}$ up until $\Sigma_{0}$ and then retraces the evolution back to the initial state instead of evolving forward from $\Sigma_{0}$ (to the future Cauchy slice $\Sigma_{T}$). Then the cut along the subregion $\mathcal{A}$ in the path integral that computes the reduced density matrix $\rho_{\mathcal{A}}$ occurs on the Cauchy slice $\Sigma_{0}$ where the forward ($-T\rightarrow 0$) and backward ($0\rightarrow -T$) branches join, as shown in figure \ref{Srho}. Under this forward-backward construction, the generating functional related to the CHEE then turns out to take the form $e^{i (S^{for}-S^{bac})}$, such that only the imaginary part of the (gravitational) action contributes. For detailed discussions in such context, refer to \cite{Colin-Ellerin:2020mva,Colin-Ellerin:2021jev}, and see also the Appendix A in \cite{Dong:2016hjy}.

However, in our real-time prescription for pseudo entropy, the path integral of $\tau^{1\vert 2}_{\mathcal{A}}$ or $\tau^{2\vert 1}_{\mathcal{A}}$ traces the entire history of the evolution, forward or backward, and then shall take the form $e^{i S^{for}}$ or $e^{-i S^{bac}}$, which ensures that the pseudo entropy shall generally take complex values. Indeed, this central feature of pseudo entropy stems from the fact that both the information of initial and final state are involved, which is inherent in the definition of the transition matrix.

\section{Complex-valued holographic pseudo entropy}\label{sec_4}
Having the real-time QFT path integral of pseudo entropy at hand, we turn to its holographic implications, and we will only briefly discuss how the complex-valued holographic pseudo entropy (HPE) comes from the real-time AdS/CFT duality, leaving the rigorous considerations as the future problem. The methodology used below is usual, refer to \emph{e.g.} \cite{Rangamani:2016dms,Nishioka:2018khk} for the reviews and the references therein.

Going to the AdS/CFT context, we assume a holographic Lorentzian-signature field theory with a semi-classical Einstein-Hilbert gravitational dual in Lorentzian AdS spacetime, and adopt the Swhinger-Keldysh time-folded geometry $\mathcal{B}$ as the boundary geometry that provides the asymptotic boundary conditions for the bulk gravity dual $\mathcal{M}$. We refer to the bulk Cauchy slices anchored on the boundary Cauchy slice $\Sigma_t$ as $\widetilde{\Sigma}_t$, and $\partial \widetilde{\Sigma}_t=\Sigma_t$. In our real-time prescription, the pseudo entropies $S(\tau^{i\vert j}_{\mathcal{A}})$ are computed separately in a manner of singling out a spacetime branch $\mathcal{B}^{i\vert j}$ with a cut along $\mathcal{A}$, and constructing the replica manifold $\mathcal{B}^{i\vert j}_{n}$, as discussed in Section 2. It would thus be dual to the gravitational problem with both the initial and final condition specified. We refer to the bulk dual to $\mathcal{B}^{i\vert j}_n$ as the covering space geometry $\mathcal{M}^{i\vert j}_n$, provided $\partial \mathcal{M}^{i\vert j}_n=\mathcal{B}^{i\vert j}_n$. As usual, we assume the boundary replica $\mathbb{Z}_n$ symmetry extends to the bulk $\mathcal{M}^{i\vert j}_n$, with $\partial \mathcal{A}$ extending naturally to a bulk codimension-2 spacelike surface $\mathbf{e}_n$, the bulk fixed point set of the $\mathbb{Z}_n$ action.

For our purpose, we will take a replica $\mathbb{Z}_n$ quotient and restrict ourselves into the the orbifold spacetime $\hat{\mathcal{M}}^{i\vert j}_n=\mathcal{M}^{i\vert j}_n/\mathbb{Z}_n$ which has the same boundary condition $\mathcal{B}^{i\vert j}$ as the original manifold $\mathcal{M}^{i\vert j}$, but with a conical singularity at the fixed locus $\mathbf{e}_n$. In this way, $\hat{\mathcal{M}}^{1\vert 2}$ and $\hat{\mathcal{M}}^{2\vert 1}$ can be thought of as the time-folded $\mathcal{M}$ with a certain conical singularity localized in its forward and backward branch, respectively. This is perhaps the most straightforward way to exploit the CPT-conjugation symmetry of the Schwinger-Keldysh geometry $\mathcal{B}$, which shall be assumed to extend into the bulk dual $\mathcal{M}$, to reveal the complex nature of HPE. Via the AdS/CFT duality, the pseudo entropies evaluated in the bulk are given as\footnote{Usually, there is an  unavoidable ambiguity when taking the logarithm of $e^{i S}$ if $i S$ is complex-valued,  \emph{i.e.}, $\log{e^{iS}}=i S + 2k\pi i$, $k\in \mathbb{Z}$. Thus the choice of the branch of $\log$ function may be important, as already discussed in \cite{PhysRevD.103.026005} for the generic pseudo entropy. This subtlety of pseudo entropy is indeed in line, and can be understood here, with the fact that the real part of $S$ only contributes a pure phase (that is generally dynamical). In this note, we simply adopt the intuitive convention: $\log{e^{iS}}\equiv i S$. That is, the branch of log here is automatically fixed by the \emph{physical} action, $S$, regarding another choice, $e.g.$, $-\pi<\textrm{Im}(\log e^{iS})\leq\pi$, as also a possibility for future considerations.}
\begin{equation}\label{bulkrenyi}
	S(\tau^{i\vert j}_{\mathcal{A}})=\lim_{n\rightarrow1}\frac{n}{1-n}\left(i S[\hat{\mathcal{M}}^{i\vert j}_n]-i S[\mathcal{M}^{i\vert j}]\right),
\end{equation}
where $S[\hat{\mathcal{M}}^{i\vert j}_n]$ is the gravitational action of the bulk geometry $\hat{\mathcal{M}}^{i\vert j}_n$ in the stationary phase approximation, and we have used the definition $n S[\hat{\mathcal{M}}_n^{i\vert j}]=S[\mathcal{M}_n^{i\vert j}]$. Viewing $\mathcal{M}^{i\vert j}$ as forward and backward segment of the time-folded bulk geometry $\mathcal{M}$ which is dual to the boundary Schwinger-Keldysh geometry, we then rewrite $S[\hat{\mathcal{M}}^{1\vert 2}]\equiv S[\mathcal{M}^{for}]$ and $S[\hat{\mathcal{M}}^{2\vert 1}]\equiv -S[\mathcal{M}^{bac}]$, noting that there contains now a contribution from the conical singularity at the locus $\mathbf{e}_n$, the codimension-2 surface homologous to the entangling surface $\partial\mathcal{A}$. The relative sign comes from the fact the $\mathcal{M}^{for}$ and $\mathcal{M}^{bac}$ are related to each other with a time reversal, and is fixed in accord with the Schwinger-Keldysh generating functional that is of the form $e^{i (S^{for}-S^{bac})}$. Assuming the CPT-conjugation symmetry extends to the bulk geometry (the bulk metric shall also be conjugated if it is complex\cite{Colin-Ellerin:2020mva}), we thus expect $S[\mathcal{M}^{bac}]=S[\mathcal{M}^{for}]^*$. Then at this stage, we can deduce that the HPE computed in the bulk indeed takes complex value (at least, it is generally complex-valued by definition), as naturally as the generic non-Hermitian transition matrix is defined.

To make above argument more apparent, let us compare the computations of the holographic pseudo entropy (HPE) and the covariant holographic entanglement entropy (CHEE).\footnote{The first notion is of course that, CHEE involves the generic time-dependent density matrices, while the computation of HPE here is just formulated into the Lorentzian path integral representation, complementary to its Euclidean counterpart, which simply generalizes the (Lorentzian) path integral of the holographic entanglement entropy (HEE) to the non-static situations.} As aforementioned, the crucial observation is that CHEE involves only the causal past of the bulk Cauchy slice $\widetilde{\Sigma}_0$ of interest ($\partial\widetilde{\Sigma}_T=\Sigma_T$) and possesses the CPT-conjugation symmetry which renders the entropies real-valued. To see this, we recall that the bulk on-shell action related to CHEE is given as,
\begin{equation}
S[\mathcal{M}]=S[\mathcal{M}^{for}]-S[\mathcal{M}^{bac}]=2 i \;\textrm{Im}(S[\mathcal{M}^{for}]),
\end{equation}
which is purely imaginary due to the pairwise cancellation between the forward and backward segments and finally gives rise to the real-valued entropies\cite{Colin-Ellerin:2021jev}. Such the pairwise cancellation may not generally happen in the case of HPE (due to the violation of the time-translation and -reflection symmetry\footnote{If the time-translation or -reflection symmetry is restored, we expect that the HPE reduces to the ordinary holographic entanglement entropy (HEE) which is real-valued.}), since the computation there involves $S[\mathcal{M}^{for}]$ or $S[\mathcal{M}^{bac}]$ only. More precisely, in HPE, the fixed point locus $\mathbf{e}_n$ is localized in the interior of the forward or backward branch of the time-folded bulk manifold, while in CHEE, $\mathbf{e}_n$ lies at the boundary between the forward and backward branches. And in the case of HPE, the extremal surface condition\cite{Hubeny:2007xt} for surface $\mathbf{e}_n$ in the $n\rightarrow 1$ limit would be recovered by examining the local geometry near this surface through the ordinary argument\cite{Dong:2016hjy}. Indeed, in the case of CHEE, before gluing the forward and backward branches across the Cauchy slice $\widetilde{\Sigma}_{0}$ to compute $\textrm{tr}_{\mathcal{A}}\rho_\mathcal{A}^n$ (near $n\sim1$), $\widetilde{\Sigma}_0$ is not arbitrary: it should pass through the extremal surface\cite{Dong:2016hjy}. To simplify our discussion on HPE, we simply continue the bulk evolution from the Cauchy slice $\widetilde{\Sigma}_0$, in CHEE, to the future Cauchy slice $\widetilde{\Sigma}_T$ ($\partial\widetilde{\Sigma}_T=\Sigma_T$) in both forward and backward branches with final condition specified on $\widetilde{\Sigma}_{T}$, then glue up the the forward and backward branches along $\widetilde{\Sigma}_{T}$ in a certain continuous manner, so that the evolution is reversed on $\widetilde{\Sigma}_T$ and we do not have boundary terms there, thus allowing us to take the time-folded bulk geometry of the Schwinger-Keldysh type to perform the computation. And recall that, this time the locus $\mathbf{e}_n$ is localized in the interior of the forward or backward branch.

Knowing $\mathbf{e}_n$ reduces to the extremal surface in the $n\rightarrow 1$ limit, we turn to evaluate the on-shell action using the Einstein gravity with conical singularity at $\mathbf{e}_n$. The evaluation turns out to be similar to that of CHEE \cite{Dong:2016hjy}, and gives,
\begin{equation}\label{final}
\begin{split}
 &S(\tau^{1\vert 2}_{\mathcal{A}})=-i \lim_{n\rightarrow1}\partial_n S[\mathcal{M}^{for}];\\
 &S[\mathcal{M}^{for}]=-\lim_{\epsilon\rightarrow 0}\frac{1}{8\pi G_N}\int_{\mathbf{e}_n(\epsilon)}\mathcal{K}^{for}_\epsilon,
\end{split}
\end{equation}
where we have regulated the codimension-2 locus $\mathbf{e}_n$ to be the codimension-1 surface $\mathbf{e}_n(\epsilon)$ defined as the hypersurface $r=\epsilon$ in the local coordinate around $\mathbf{e}_n$, with $\mathcal{K}^{for}_{\epsilon}$ being the trace of the extrinsic curvature of $\mathbf{e}_n(\epsilon)$ located in the forward branch. And the definition of \emph{modular entropy} is used in the computation\cite{Dong:2016hjy}. According to the previous discussion, the pseudo entropy $S(\tau^{2\vert1}_{\mathcal{A}})$ evaluated in the backward branch is expected to be obtained as the complex conjugate of $S(\tau^{1\vert2}_{\mathcal{A}})$. We restrict ourselves to the case of the real-valued bulk metric. As in the CHEE, the light-like singularities of the extrinsic curvature contribute the imaginary part of the action and then the real part of entropy, which formally ends up with the HRT formula\cite{Hubeny:2007xt}. However, there is also additional imaginary contribution to entropy from the integral of regular extrinsic curvature, as long as it survives in the $\epsilon\rightarrow 0$ limit, which is now understood as the imaginary part of the pseudo entropy. In this note, we refer to these contributions from the integral of the extrinsic curvature of the regularized conical singularity locus $\mathbf{e}_n(\epsilon)$ in the $n\rightarrow 1$ limit as the regularized extrinsic curvature term of the extremal surface.\footnote{We expect this regularized extrinsic curvature term turns out to be exactly the Gibbons-Hawking boundary term of the regularized extremal surface $\mathcal{E}$, namely $S(\tau_{\mathcal{A}}^{1\vert2})=i\frac{1}{8\pi G_N}\lim_{\epsilon\rightarrow 0}\int_{\mathcal{E}(\epsilon)} \mathcal{K}^{for}_\epsilon$, where $\mathcal{E}(\epsilon)\equiv \mathbf{e}_n(\epsilon)\vert_{n\rightarrow 1}$ is just the regularized extremal surface. Instead of starting computing \eqref{final} with brute force, one can naively infer this in an alternative way as follows. Recalling that the solution to the Einstein equation in the covering space is smooth and one introduces the artificial singularity when taking the quotient space, \emph{i.e.} defining $S[\hat{\mathcal{M}}_n]=S[\mathcal{M}_n]/n$. Thus, rather than regularizing the artificial singularity in $S[\hat{\mathcal{M}}_n]$, we can alternatively regulate the $\mathbb{Z}_n$ fixed point locus $\mathbf{e}_n$ to $\mathbf{e}_n(\epsilon)$ in the covering space and then we take the quotient. This can be achieved by adding the imaginary boundary terms of $\mathbf{e}_n(\epsilon)$ to $S[\mathcal{M}_n]$, \emph{i.e.}, $S[\mathcal{M}_n]=S[\mathcal{M}_n]+S_{bdy}[\mathbf{e}_n(\epsilon)]-S_{bdy}[\mathbf{e}_n(\epsilon)]$, then after rescaling the period of the modular time (if explicitly defined) we suggest that $S[\hat{ \mathcal{M}}_n]=S[\mathcal{M}_n]^\prime+(1-\frac{1}{n})S_{bdy}[\mathbf{e}_n(\epsilon)]^\prime$, where $S[\mathcal{M}_n]^\prime$ is the ordinary covering space solution that is now measured in the rescaled modular time (so is $S_{bdy}[\mathbf{e}_n(\epsilon)]^\prime$) and is still regular. The factor $\frac{1}{n}$ comes from the fact that we defined $S[\hat{\mathcal{M}}_n]=S[\mathcal{M}_n]/n$ which exactly introduces a singularity across $\mathbf{e}_n(\epsilon)$ in the quotient space. Of course, this regularization only makes sense in the $\epsilon\rightarrow 0$ limit, since the singularity is merely localized at the $\mathbb{Z}_n$ fixed locus $\mathbf{e}_n$. Finally, by taking the $n$-derivative and then the $n\rightarrow 1$ limit, we arrive at the expected result.}

The computations of HPE require the overall knowledge of the local (time-dependent) geometry around the extremal surfaces. This seems to agree with the argument that, the transition matrix involving both the initial and final state such that the causal past and the causal future of the Cauchy slice of interest and the information of the evolution shall be included to compute the pseudo entropy. And via AdS/CFT, its dual bulk on-shell action is localized on the boundary of the extremal surface, thus it is natural to expect the local geometry around the extremal surface somewhat encodes the information of the evolution with initial and final state both specified, and especially encodes the difference between the initial and final state, which should generally appear as a dynamical phase. It is then intriguing to think of the HPE as a generalization of the CHEE that picks up only the partial information of the extremal surfaces, while the CHEE and the ordinary holographic entanglement entropy (HEE) can both be thought of as involving only the initial state.
\section{Conclusion}
We provided a real-time prescription for the computations of the pseudo entropy in the quantum field theory, and recast it into the Schwinger-Keldysh formalism which makes the complex nature of the pseudo entropy more apparent. We also draw a line between the real-time path integral representations of the holographic pseudo entropy and covariant holographic entanglement entropy, and explain how the holographic pseudo entropy develops complex values through the argument of the CPT-conjugation symmetry of the Schwinger-Keldysh contour. Via the real-time AdS/CFT correspondence, we argue that the holographic pseudo entropy is indeed generally complex-valued and may be dual to the bulk codimension-2 extremal surface receiving the complex contribution from the regularized extrinsic curvature term of the extremal surface, which can be considered as generalizing the covariant holographic entanglement entropy. It can be believed that, the holographic pseudo entropy stands as an inherent and necessary entanglement measure corresponding to the problems of the dynamical gravity. We also hope this work can to some extent shed light on the dynamical aspect of the AdS/CFT correspondence and finally the quantum gravity.
\section*{Acknowledgements}
This work is not supported by any funding.
\appendix
\bibliographystyle{JHEP}
\bibliography{rtHPEref}
\appendix
\end{document}